\documentclass[12pt]{article}

\usepackage{epsf}
\usepackage{rotate}

\newcommand{\lsim}{
\mathrel{\hbox{\rlap{\hbox{\lower4pt\hbox{$\sim$}}}\hbox{$<$}}}}
\newcommand{\gsim}{
\mathrel{\hbox{\rlap{\hbox{\lower4pt\hbox{$\sim$}}}\hbox{$>$}}}}

\begin{document}


\begin{titlepage}

\thispagestyle{empty}

\begin{flushright}
CERN-TH/2003-241\\
hep-ph/0310081
\end{flushright}

\vspace{2.0truecm}
\begin{center}
\boldmath
\large\bf New Strategies to Extract Weak Phases from Neutral B Decays
\unboldmath
\end{center}

\vspace{1.2cm}
\begin{center}
Robert Fleischer\\[0.1cm]
{\sl Theory Division, CERN, CH-1211 Geneva 23, Switzerland}
\end{center}

\vspace{1.7cm}
\begin{abstract}
\vspace{0.2cm}\noindent
We discuss new, theoretically clean strategies to determine the 
angle $\gamma$ of the unitarity triangle from $B_d\to DK_{\rm S(L)}$, 
$B_s\to D\eta^{(')}, D\phi, ...$\ decays, and point out that the
$B_s\to DK_{\rm S(L)}$ and $B_d\to D\pi^0, D\rho^0, ...$\ modes 
allow very interesting determinations of the $B^0_q$--$\overline{B^0_q}$
mixing phases $\phi_s$ and $\phi_d$, respectively. Their colour-allowed
counterparts $B_s\to D_s^{(\ast)\pm} K^\mp, ...$\ and 
$B_d\to D^{(\ast)\pm} \pi^\mp, ...$\ also offer new methods to probe $\gamma$.
\end{abstract}

\vspace{1.8truecm}

\begin{center}
{\sl Invited talk at the HEP2003 Europhysics Conference\\
Aachen, Germany, 17--23 July 2003\\
To appear in the Proceedings}
\end{center}

\vfill
\noindent
CERN-TH/2003-241\\
October 2003

\newpage
\thispagestyle{empty}
\vbox{}
\newpage
 
\setcounter{page}{1}

\end{titlepage}


%
%
%
\section{Introduction}\label{intro}
The time-dependent CP asymmetries of neutral $B_q$-meson
decays ($q\in\{d,s\}$) into CP eigenstates, which satisfy  
$({\cal CP})|f\rangle=\pm|f\rangle$, provide valuable information
\cite{RF-rev}:
\begin{eqnarray}
\lefteqn{
\frac{\Gamma(B^0_q(t)\to f)-
\Gamma(\overline{B^0_q}(t)\to f)}{\Gamma(B^0_q(t)\to f)+
\Gamma(\overline{B^0_q}(t)\to f)}}\nonumber\\
&&=\frac{{\cal A}_{\rm CP}^{\rm dir}\cos(\Delta M_q t)+
{\cal A}_{\rm CP}^{\rm mix}\sin(\Delta M_q t)}{\cosh(\Delta\Gamma_qt/2)-
{\cal A}_{\rm \Delta\Gamma}\sinh(\Delta\Gamma_qt/2)}.
\end{eqnarray}
Here the CP-violating observables 
\begin{equation}\label{obs}
{\cal A}^{\mbox{{\scriptsize dir}}}_{\mbox{{\scriptsize CP}}}
\equiv\frac{1-\bigl|\xi_f^{(q)}\bigr|^2}{1+
\bigl|\xi_f^{(q)}\bigr|^2} \quad \mbox{and} \quad
{\cal A}^{\mbox{{\scriptsize mix}}}_{\mbox{{\scriptsize
CP}}}\equiv\frac{2\,\mbox{Im}\,\xi^{(q)}_f}{1+\bigl|\xi^{(q)}_f\bigr|^2}
\end{equation}
originate from ``direct'' and ``mixing-induced'' CP violation, 
respectively, and are governed by 
\begin{equation}
\xi_f^{(q)}=-e^{-i\phi_q}
\left[\frac{A(\overline{B_q^0}\to f)}{A(B_q^0\to f)}\right],
\end{equation}
where
\begin{equation}
\phi_q\stackrel{{\rm SM}}{=}2\,\mbox{arg}(V_{tq}^\ast V_{tb})
=\left\{\begin{array}{cc}
+2\beta & \mbox{($q=d$)}\\
-2\lambda^2\eta & 
\mbox{($q=s$)}
\end{array}\right.
\end{equation}
is the CP-violating weak $B^0_q$--$\overline{B^0_q}$ mixing phase. The width 
difference $\Delta\Gamma_q$, which may be sizeable in the $q=s$ case, 
offers another
observable ${\cal A}_{\rm \Delta\Gamma}$, which is, however, not independent
from those in (\ref{obs}), and can be extracted from the following
``untagged'' rates:
\begin{eqnarray}
\lefteqn{\langle\Gamma(B_q(t)\to f)\rangle\equiv
\Gamma(B_q^0(t)\to f)+\Gamma(\overline{B^0_q}(t)\to f)}\nonumber\\
&&\propto\left[\cosh(\Delta\Gamma_qt/2)-{\cal A}_{\Delta\Gamma}
\sinh(\Delta\Gamma_qt/2)\right]e^{-\Gamma_qt}.\label{un-tagged-def}
\end{eqnarray}

\boldmath
\section{$B_d\to DK_{\rm S(L)}$, 
$B_s\to D\eta^{(')}, D\phi, ...$ and $B_s\to DK_{\rm S(L)}$,
$B_d\to D\pi^0, D\rho^0, ...$}\label{partI}
\unboldmath
Let us consider in this section $B^0_q\to D^0f_r$ transitions, where 
$r\in\{s,d\}$ distinguishes between $b\to Ds$ and $b\to D d$ 
processes \cite{CS-PLB,CS-NPB}. If we require 
$({\cal CP})|f_r\rangle=\eta_{\rm CP}^{f_r}|f_r\rangle$,
$B^0_q$ and $\overline{B^0_q}$ mesons may both decay into $D^0f_r$, 
thereby leading to interference effects between $B^0_q$--$\overline{B^0_q}$ 
mixing and decay processes, which involve the weak phase $\phi_q+\gamma$: 
\begin{itemize}
\item For $r=s$, i.e.\ $B_d\to D K_{\rm S(L)}$, 
$B_s\to D\eta^{(')}, D\phi$, ..., these effects are governed by a hadronic
parameter $x_{f_s}e^{i\delta_{f_s}}\propto R_b\approx0.4$, and are hence 
favourably large. 
\item For $r=d$, i.e.\ $B_s\to DK_{\rm S(L)}$, $B_d\to D\pi^0, D\rho^0$ ..., 
these effects are tiny because of 
$x_{f_d}e^{i\delta_{f_d}}\propto -\lambda^2R_b\approx -0.02$. 
\end{itemize}

\boldmath
\subsection{$B_d\to DK_{\rm S(L)}$, 
$B_s\to D\eta^{(')}, D\phi, ...$}
\unboldmath
Let us first focus on $r=s$. If we make use of the CP eigenstates
$D_\pm$ of the neutral $D$-meson system satisfying 
$({\cal CP})|D_\pm\rangle=\pm|D_\pm\rangle$, we obtain 
additional interference effects between $B_q^0\to D^0 f_s$ and 
$B_q^0\to \overline{D^0} f_s$ at the decay-amplitude level, which involve 
$\gamma$. The most straightforward observable we may measure is the 
``untagged'' rate
\begin{eqnarray}
\lefteqn{\langle\Gamma(B_q(t)\to D_\pm f_s)\rangle\equiv}\nonumber\\
&&\Gamma(B^0_q(t)\to D_\pm f_s)+
\Gamma(\overline{B^0_q}(t)\to D_\pm f_s)\nonumber\\
&&\stackrel{\Delta\Gamma_q=0}{=}
\left[\Gamma(B^0_q\to D_\pm f_s)+
\Gamma(\overline{B^0_q}\to D_\pm f_s)\right]e^{-\Gamma_qt}\nonumber\\
&&\mbox{}~~~\equiv \langle\Gamma(B_q\to D_\pm f_s)\rangle e^{-\Gamma_qt},
\end{eqnarray}
providing the following ``untagged'' rate asymmetry:
\begin{equation}
\Gamma_{+-}^{f_s}\equiv
\frac{\langle\Gamma(B_q\to D_+ f_s)\rangle-\langle
\Gamma(B_q\to D_- f_s)\rangle}{\langle\Gamma(B_q\to D_+ f_s)\rangle
+\langle\Gamma(B_q\to D_- f_s)\rangle}.
\end{equation}
Interestingly, already this quantity offers valuable information on
$\gamma$, since bounds on this angle are implied by
\begin{equation}
|\cos\gamma|\geq |\Gamma_{+-}^{f_s}|. 
\end{equation}
Moreover, if we fix the sign of $\cos\delta_{f_s}$ with the help of 
the factorization approach, we obtain 
\begin{equation}\label{sgn-cos-gam}
\mbox{sgn}(\cos\gamma)=-\mbox{sgn}(\Gamma_{+-}^{f_s}),
\end{equation} 
i.e.\ we may decide whether $\gamma$ is smaller or larger than 
$90^\circ$.
If we employ, in addition, the mixing-induced observables 
$S_\pm^{f_s}\equiv {\cal A}_{\rm CP}^{\rm mix}(B_q\to D_\pm f_s)$, 
we may determine $\gamma$. To this end, it is convenient to introduce
the quantities
\begin{equation}
\langle S_{f_s}\rangle_\pm\equiv\frac{S_+^{f_s}\pm S_-^{f_s}}{2}.
\end{equation}
Expressing the $\langle S_{f_s}\rangle_\pm$ in terms of the $B_q\to D_\pm f_s$ 
decay parameters gives rather complicated formulae. However, 
complementing the $\langle S_{f_s}\rangle_\pm$ with $\Gamma_{+-}^{f_s}$ 
yields 
\begin{equation}\label{key-rel}
\tan\gamma\cos\phi_q=
\left[\frac{\eta_{f_s} \langle S_{f_s}
\rangle_+}{\Gamma_{+-}^{f_s}}\right]+\left[\eta_{f_s}\langle S_{f_s}\rangle_--
\sin\phi_q\right],
\end{equation}
where $\eta_{f_s}\equiv(-1)^L\eta_{\rm CP}^{f_s}$, with $L$ denoting the 
$Df_s$ angular momentum \cite{CS-PLB}. If we use this simple -- but 
{\it exact} -- relation, 
we obtain the twofold solution $\gamma=\gamma_1\lor\gamma_2$, with 
$\gamma_1\in[0^\circ,180^\circ]$ and $\gamma_2=\gamma_1+180^\circ$. Since 
$\cos\gamma_1$ and $\cos\gamma_2$ have opposite signs, (\ref{sgn-cos-gam}) 
allows us to fix $\gamma$ {\it unambiguously}. Another advantage of 
(\ref{key-rel}) is that $\langle S_{f_s}\rangle_+$ and $\Gamma_{+-}^{f_s}$ 
are both proportional to $x_{f_s}\approx0.4$, so that the first term in square 
brackets is of ${\cal O}(1)$, whereas the second one is of 
${\cal O}(x_{f_s}^2)$, hence playing a minor r\^ole. In order to extract 
$\gamma$, we may also employ $D$ decays into CP non-eigenstates 
$f_{\rm NE}$, where we have to deal with complications originating from 
$D^0,\overline{D^0}\to f_{\rm NE}$ interference effects \cite{KayLo}.
Also in this case, $\Gamma_{+-}^{f_s}$ is a very powerful ingredient, 
offering an efficient, {\it analytical} strategy to include these 
interference effects in the extraction of $\gamma$ \cite{CS-NPB}.

\boldmath
\subsection{$B_s\to DK_{\rm S(L)}$,
$B_d\to D\pi^0, D\rho^0, ...$}
\unboldmath
The $r=d$ case also has interesting features. It corresponds to 
$B_s\to DK_{\rm S(L)}$, $B_d\to D\pi^0, D\rho^0$ ...\ decays, which
can be described through the same formulae as their $r=s$ counterparts. 
Since the relevant interference effects are governed by $x_{f_d}\approx -0.02$,
these channels are not as attractive for the extraction of $\gamma$ as the 
$r=s$ modes. On the other hand, the relation
\begin{equation}
\eta_{f_d}\langle S_{f_d}\rangle_-=
\sin\phi_q + {\cal O}(x_{f_d}^2)
=\sin\phi_q + {\cal O}(4\times 10^{-4})
\end{equation}
offers very interesting determinations of $\sin\phi_q$ \cite{CS-PLB}. 
Following this avenue, there are no penguin uncertainties, and the theoretical 
accuracy is one order of magnitude better than in the ``conventional'' 
$B_d\to J/\psi K_{\rm S}$, $B_s\to J/\psi \phi$ strategies. In particular,
$\phi_s^{\rm SM}=-2\lambda^2\eta$ could, in principle, be determined with 
a theoretical uncertainty of only ${\cal O}(1\%)$, in contrast to the
extraction from the $B_s\to J/\psi \phi$ angular distribution, which
suffers from generic penguin uncertainties at the $10\%$ level.

\boldmath
\section{$B_s\to D_s^{(\ast)\pm} K^\mp, ...$\ and 
$B_d\to D^{(\ast)\pm} \pi^\mp, ...$}\label{partII}
\unboldmath
Let us now consider the colour-allowed counterparts of the 
$B_q\to D f_q$ modes discussed above, which we may write generically 
as $B_q\to D_q \overline{u}_q$ \cite{CA-NPB}. The characteristic feature 
of these transitions is that both a $B^0_q$ and a $\overline{B^0_q}$ meson 
may decay into $D_q \overline{u}_q$, thereby leading to interference between 
$B^0_q$--$\overline{B^0_q}$ mixing and decay processes, which involve the 
weak phase $\phi_q+\gamma$:
\begin{itemize}
\item In the case of $q=s$, i.e.\ 
$D_s\in\{D_s^+, D_s^{\ast+}, ...\}$ and $u_s\in\{K^+, K^{\ast+}, ...\}$, 
these effects are favourably large as they are governed by 
$x_s e^{i\delta_s}\propto R_b\approx0.4$.
\item In the case of $q=d$, i.e.\ $D_d\in\{D^+, D^{\ast+}, ...\}$ 
and $u_d\in\{\pi^+, \rho^+, ...\}$, the interference effects are described 
by $x_d e^{i\delta_d}\propto -\lambda^2R_b\approx-0.02$, and hence are tiny. 
\end{itemize}
We shall only consider $B_q\to D_q \overline{u}_q$ modes, 
where at least one of the $D_q$, $\overline{u}_q$ states is a pseudoscalar 
meson; otherwise a complicated angular analysis has to be performed.
 
It is well known that such decays allow determinations of the weak
phases $\phi_q+\gamma$, where the ``conventional'' approach works
as follows \cite{BDpi-conv,BsDsK-conv}: if we measure the observables 
$C(B_q\to D_q\overline{u}_q)\equiv C_q$ 
and $C(B_q\to \overline{D}_q u_q)\equiv \overline{C}_q$ provided by the
$\cos(\Delta M_qt)$ pieces of the time-dependent rate asymmetries, 
we may determine $x_q$ from terms entering at the $x_q^2$ 
level. In the case of $q=s$, we have $x_s={\cal O}(R_b)$, 
implying $x_s^2={\cal O}(0.16)$, so that this may actually be possible, 
although challenging. On the other hand, 
$x_d={\cal O}(-\lambda^2R_b)$ is doubly Cabibbo-suppressed.
Although it should be possible to resolve terms of ${\cal O}(x_d)$, 
this will be impossible for the vanishingly small $x_d^2={\cal O}(0.0004)$ 
terms, so that other approaches to fix $x_d$ are required \cite{BDpi-conv}. 
In order to extract $\phi_q+\gamma$, the mixing-induced observables 
$S(B_q\to D_q\overline{u}_q)\equiv S_q$ and 
$S(B_q\to \overline{D}_q u_q)\equiv \overline{S}_q$ associated with the
$\sin(\Delta M_qt)$ terms of the time-dependent rate 
asymmetries must be measured, where it is convenient to introduce
\begin{equation}\label{Savpm}
\langle S_q\rangle_\pm\equiv
\frac{\overline{S}_q\pm S_q}{2}.
\end{equation}
If we assume that $x_q$ is known, we may consider
\begin{eqnarray}
s_+&\equiv& (-1)^L
\left[\frac{1+x_q^2}{2 x_q}\right]\langle S_q\rangle_+
=+\cos\delta_q\sin(\phi_q+\gamma)\\
s_-&\equiv&(-1)^L
\left[\frac{1+x_q^2}{2x_q}\right]\langle S_q\rangle_-
=-\sin\delta_q\cos(\phi_q+\gamma),
\end{eqnarray}
yielding
\begin{equation}\label{conv-extr}
\sin^2(\phi_q+\gamma)=\frac{1+s_+^2-s_-^2}{2} \pm
\sqrt{\frac{(1+s_+^2-s_-^2)^2-4s_+^2}{4}},
\end{equation}
which implies an eightfold solution for $\phi_q+\gamma$. If we fix the
sign of $\cos\delta_q$ with the help of factorization, a fourfold 
discrete ambiguity emerges. Note that this assumption allows us to 
extract also the sign of $\sin(\phi_q+\gamma)$ from $\langle S_q\rangle_+$,
which is of particular interest, as discussed in \cite{CA-NPB}.
To this end, the factor $(-1)^L$, where $L$ is the $D_q\overline{u}_q$ 
angular momentum, has to be properly taken into account.

Let us now discuss the new strategies to explore the 
$B_q\to D_q \overline{u}_q$ modes proposed in \cite{CA-NPB}. If 
$\Delta\Gamma_s$ is sizeable, the time-dependent ``untagged'' rates
introduced in (\ref{un-tagged-def}) 
\begin{eqnarray}
\lefteqn{\langle\Gamma(B_q(t)\to D_q\overline{u}_q)\rangle=
\langle\Gamma(B_q\to D_q\overline{u}_q)\rangle 
e^{-\Gamma_qt}}\label{untagged}\\
&&\times\left[\cosh(\Delta\Gamma_qt/2)-{\cal A}_{\rm \Delta\Gamma}
(B_q\to D_q\overline{u}_q)\,\sinh(\Delta\Gamma_qt/2)\right]\nonumber
\end{eqnarray}
and their CP conjugates provide 
${\cal A}_{\rm \Delta\Gamma}(B_s\to D_s\overline{u}_s)
\equiv {\cal A}_{\rm \Delta\Gamma_s}$ and 
${\cal A}_{\rm \Delta\Gamma}(B_s\to \overline{D}_s u_s)\equiv 
\overline{{\cal A}}_{\rm \Delta\Gamma_s}$, which yield 
\begin{equation}\label{untagged-extr}
\tan(\phi_s+\gamma)=
-\left[\frac{\langle S_s\rangle_+}{\langle{\cal A}_{\rm \Delta\Gamma_s}
\rangle_+}\right]
=+\left[\frac{\langle{\cal A}_{\rm \Delta\Gamma_s}
\rangle_-}{\langle S_s\rangle_-}\right],
\end{equation}
where the $\langle{\cal A}_{\rm \Delta\Gamma_s}\rangle_\pm$ are defined
in analogy to (\ref{Savpm}). These relations allow an 
{\it unambiguous} extraction of $\phi_s+\gamma$ if we fix again the sign
of $\cos\delta_q$ through factorization. Another important
advantage of (\ref{untagged-extr}) is that we do {\it not} have to rely on 
${\cal O}(x_s^2)$ terms, as $\langle S_s\rangle_\pm$ and 
$\langle {\cal A}_{\rm \Delta\Gamma_s}\rangle_\pm$ are proportional to $x_s$. 
On the other hand, we need a sizeable value of $\Delta\Gamma_s$. Measurements 
of untagged rates are also very useful in the case of vanishingly small 
$\Delta\Gamma_q$, since the ``unevolved'' untagged rates 
in (\ref{untagged}) offer various interesting strategies to determine 
$x_q$ from the ratio of $\langle\Gamma(B_q\to D_q\overline{u}_q)\rangle+
\langle\Gamma(B_q\to \overline{D}_q u_q)\rangle$ to CP-averaged rates of
appropriate $B^\pm$ or flavour-specific $B_q$ decays.

If we keep the hadronic parameter $x_q$ and the associated strong phase
$\delta_q$ as ``unknown'', free parameters in the expressions for the
$\langle S_q\rangle_\pm$, we obtain 
\begin{equation}
|\sin(\phi_q+\gamma)|\geq|\langle S_q\rangle_+|, \quad
|\cos(\phi_q+\gamma)|\geq|\langle S_q\rangle_-|,
\end{equation}
which can straightforwardly be converted into bounds on $\phi_q+\gamma$. 
If $x_q$ is known, stronger constraints are implied by 
\begin{equation}\label{bounds}
|\sin(\phi_q+\gamma)|\geq|s_+|, \quad
|\cos(\phi_q+\gamma)|\geq|s_-|.
\end{equation}
Once $s_+$ and $s_-$ are known, we may of course determine
$\phi_q+\gamma$ through the ``conventional'' approach, using 
(\ref{conv-extr}). However, the bounds following from (\ref{bounds})
provide essentially the same information and are much simpler to 
implement. Moreover, as discussed in detail in \cite{CA-NPB}
for several examples, the bounds following from the $B_s$ and $B_d$ 
modes may be highly complementary, thereby providing particularly narrow, 
theoretically clean ranges for $\gamma$. 

Let us now further exploit the complementarity between the
$B_s^0\to D_s^{(\ast)+}K^-$ and $B_d^0\to D^{(\ast)+}\pi^-$ modes.
If we look at their decay topologies, we observe that these channels are 
related to each other through an interchange of all down and strange quarks. 
Consequently, the $U$-spin flavour symmetry of strong interactions implies 
$a_s=a_d$ and $\delta_s=\delta_d$, where $a_s=x_s/R_b$ and 
$a_d=-x_d/(\lambda^2 R_b)$ are the ratios of hadronic matrix elements 
entering $x_s$ and $x_d$, respectively. There are various possibilities 
to implement these relations. A particularly simple picture emerges if we 
assume that $a_s=a_d$ {\it and} $\delta_s=\delta_d$, 
which yields
\begin{equation}
\tan\gamma=-\left[\frac{\sin\phi_d-S
\sin\phi_s}{\cos\phi_d-S\cos\phi_s}
\right]\stackrel{\phi_s=0^\circ}{=}
-\left[\frac{\sin\phi_d}{\cos\phi_d-S}\right].
\end{equation}
Here we have introduced
\begin{equation}
S=-R\left[\frac{\langle S_d\rangle_+}{\langle S_s\rangle_+}\right]
\end{equation}
with
\begin{equation}
R= \left(\frac{1-\lambda^2}{\lambda^2}\right)
\left[\frac{1}{1+x_s^2}\right],
\end{equation}
which can be fixed from untagged $B_s$ rates through
\begin{equation}
R=\left(\frac{f_K}{f_\pi}\right)^2
\left[\frac{\Gamma(\overline{B^0_s}\to D_s^{(\ast)+}\pi^-)+
\Gamma(B^0_s\to D_s^{(\ast)-}\pi^+)}{\langle\Gamma(B_s\to D_s^{(\ast)+}K^-)
\rangle+\langle\Gamma(B_s\to D_s^{(\ast)-}K^+)\rangle}\right].
\end{equation}
Alternatively, we may {\it only} assume that $\delta_s=\delta_d$ {\it or} 
that $a_s=a_d$. Apart from features related to multiple discrete ambiguities, 
the most important advantage with respect to the ``conventional'' approach 
is that the experimental resolution of the $x_q^2$ terms is not required. In 
particular, $x_d$ does {\it not} have to be fixed, and $x_s$ may only enter 
through a $1+x_s^2$ correction, which can straightforwardly be determined 
through untagged $B_s$ rate measurements. In the most refined implementation 
of this strategy, the measurement of $x_d/x_s$ would {\it only} be interesting 
for the inclusion of $U$-spin-breaking effects in $a_d/a_s$. Moreover, we may 
obtain interesting insights into hadron dynamics and $U$-spin-breaking 
effects.

\section{Conclusions}\label{concl}
We have discussed new strategies to explore CP violation through 
neutral $B_q$ decays. In the first part, we have shown that 
$B_d\to D K_{\rm S(L)}$, $B_s\to D\eta^{(')}, D\phi$, ...\ modes 
provide theoretically clean, efficient and unambiguous extractions of 
$\tan\gamma$ if we combine an ``untagged'' rate asymmetry with 
mixing-induced observables. 
On the other hand, their $B_s\to D_\pm K_{\rm S(L)}$, 
$B_d\to D_\pm \pi^0, D_\pm \rho^0$, ...\ counterparts are not as 
attractive for the determination of $\gamma$, but allow extremely
clean extractions of the mixing phases $\phi_s$ and $\phi_d$, which may be 
particularly interesting for the $\phi_s$ case. In the second part, 
we have discussed interesting new aspects of 
$B_s\to D_s^{(\ast)\pm} K^\mp$, ...\ and 
$B_d\to D^{(\ast)\pm} \pi^\mp$, ...\ decays. The observables of these modes
provide clean bounds on $\phi_q+\gamma$, where the resulting ranges for 
$\gamma$ may be highly complementary in the $B_s$ and $B_d$ cases, thereby 
yielding stringent constraints on $\gamma$. Moreover, it is of great 
advantage to combine the $B_d\to D^{(\ast)\pm} \pi^\mp$ modes with their 
$U$-spin counterparts $B_s\to D_s^{(\ast)\pm} K^\mp$, allowing us to 
overcome the main problems of the ``conventional'' strategies to deal with 
these modes. 
We strongly encourage detailed feasibility studies of these new strategies.


\begin{thebibliography}{99}
%
%
\bibitem{RF-rev}R. Fleischer,
Phys.\ Rep.\ {\bf 370} (2002) 537.

\bibitem{CS-PLB}R. Fleischer,
Phys.\ Lett.\ B {\bf 562} (2003) 234.

\bibitem{CS-NPB}R. Fleischer,
Nucl.\ Phys.\ B {\bf 659} (2003) 321.

\bibitem{KayLo}B. Kayser and D. London,
Phys.\ Rev.\ D {\bf 61} (2000) 116013.

\bibitem{CA-NPB}R. Fleischer,
hep-ph/0304027, to appear in Nucl.\ Phys.\ B.

\bibitem{BDpi-conv}I. Dunietz and R.G. Sachs,
Phys.\ Rev.\ D {\bf 37} (1988) 3186 [E: D {\bf 39} (1989) 3515];
I. Dunietz,
Phys.\ Lett.\ B {\bf 427} (1998) 179;
D.A. Suprun {\it et al.}, 
Phys.\ Rev.\ D {\bf 65} (2002) 054025.

\bibitem{BsDsK-conv}R. Aleksan {\it et al.},
Z.\ Phys.\ C {\bf 54} (1992) 653.


\end{thebibliography}
\end{document}